\documentstyle[pra,aps,eqsecnum,multicol,epsf]{revtex}

\begin{document}

\title{Griffiths-McCoy Singularities in the 
Random Transverse-Field Ising Spin Chain}

\author{Ferenc Igl\'oi$^{1,2}$, R\'obert Juh\'asz$^{2,1}$ 
and Heiko Rieger$^{3,4}$}

\address{
$^1$ Research Institute for Solid State Physics and Optics, 
H-1525 Budapest, P.O.Box 49, Hungary\\
$^2$ Institute for Theoretical Physics,
Szeged University, H-6720 Szeged, Hungary\\
$^3$ Institut f\"ur Theoretische Physik, Universit\"at zu K\"oln, 
     50923 K\"oln, Germany\\
$^4$ NIC c/o Forschungszentrum J\"ulich, 52425 J\"ulich, Germany\\
}

\date{November 11, 1998}

\maketitle

\begin{abstract}
  We consider the paramagnetic phase of the random transverse-field
  Ising spin chain and study the dynamical properties by numerical
  methods and scaling considerations. We extend our previous work
  [Phys. Rev. B {\bf 57}, 11404 (1998)] to new quantities, such as the
  non-linear susceptibility, higher excitations and the energy-density
  autocorrelation function. We show that in the Griffiths phase all
  the above quantities exhibit power-law singularities and the
  corresponding critical exponents, which vary with the distance from
  the critical point, can be related to the dynamical exponent $z$,
  the latter being the positive root of
  $\left[(J/h)^{1/z}\right]_{\rm av}=1$. Particularly, whereas the average
  spin autocorrelation function in imaginary time decays as
  $[G]_{\rm av}(\tau) \sim \tau^{-1/z}$, the average energy-density
  autocorrelations decay with another exponent as $[G^e]_{\rm av}(\tau)
  \sim \tau^{-2-1/z}$.
\end{abstract}


\newcommand{\bc}{\begin{center}}
\newcommand{\ec}{\end{center}}
\newcommand{\be}{\begin{equation}}
\newcommand{\ee}{\end{equation}}
\newcommand{\beqn}{\begin{eqnarray}}
\newcommand{\eeqn}{\end{eqnarray}}

\begin{multicols}{2}
\narrowtext

\section{Introduction}

Quantum phase transitions occur at zero temperature by varying a
parameter of the Hamiltonian, e.g. the strength of a transverse
field. Quenched, i.e. time independent disorder has generally a
profound effect on the properties of the quantum system not only at
the critical point, but also in a whole region, which extends in
both sides of the critical point. In this so called Griffiths phase
the dynamical properties of the random quantum systems are exceptional:
for example the (imaginary) time dependent average
spin-spin correlations decay algebraically\cite{qsg}
\be
[G]_{\rm av}(\tau) \sim \tau^{-1/z(\delta)}\;,
\label{auto}
\ee
where the dynamical exponent $z(\delta)$ is a continuous function of
the quantum control parameter, $\delta$. From here on we use
$[\dots]_{\rm av}$ to denote averaging over quenched disorder.  The
physical origin of this type of singular behavior, as was pointed out
by Griffiths\cite{griffiths} for classical systems, is the existence
of clusters in the random system, which are more strongly coupled than
the average. The spins of such clusters, being locally in the "ordered
phase", behave coherently as a giant spin and the corresponding
relaxation time is very large. Thus in an infinite system there is no
(finite) time-scale and, as a consequence, the autocorrelations decay
algebraically, as in (\ref{auto})\cite{classical}.

Several physical quantities, which involve an integral of the
autocorrelation function, (e.g. the static susceptibility,) are
singular not only at the critical point but also in a finite region of
the paramagnetic phase. This phenomenon has been first noticed by
McCoy in a two-dimensional classical model with correlated disorder,
(equivalent to a one-dimensional random quantum model)\cite{mccoy},
therefore we call the Griffiths singularities in quantum systems
Griffiths-McCoy singularities.

Many of the theoretical studies on random quantum systems are related
to random quantum ferromagnets\cite{fm2d} and quantum spin
glasses\cite{sglass}, which have also experimental
realizations\cite{experiment}. In higher ($d=2$ and $d=3$) dimensions
one generally studies the distribution of the (linear and non-linear)
susceptibilities, the asymptotic behavior of those can be related to
the dynamical exponent, $z(\delta)$, by scaling considerations.
According to numerical studies - in agreement with these
phenomenological theories - $z(\delta)$ is found continuous function
of the quantum control parameter $\delta$, which appear to have a
finite limiting value at the critical point, $\delta=0$, of spin
glasses\cite{sglass}, whereas it is diverging for random
ferromagnets\cite{fm2d}.

Many features of Griffiths-McCoy singularities can already be seen in
one-dimensional systems, where many exact and conjectured results
exist.  In this paper we consider the prototype of random quantum
systems the random transverse-field Ising model (RTIM) in one
dimension, defined by the Hamiltonian:
\be
H=-\sum_l J_l \sigma_l^x \sigma_{l+1}^x-\sum_l h_l \sigma_l^z\;.
\label{hamilton}
\ee
Here the $\sigma_l^x$, $\sigma_l^z$ are Pauli matrices at site $l$ and
the $J_l$ exchange couplings and the $h_l$ transverse-fields are
random variables with distributions $\pi(J)$ and $\rho(h)$,
respectively. Note that in one dimension all the couplings and fields can
be taken positive through a gauge transformation.
The model in (\ref{hamilton}) is in the ferromagnetic (paramagnetic)
phase if the couplings in average are stronger (weaker) than the transverse
fields. As a convenient quantum control-parameter one can define:
\be
\delta={[\ln h]_{\rm av}-[\ln J]_{\rm av} \over \rm{var}[\ln h]+\rm{var}[\ln J]}\;,
\label{delta}
\ee
and at the critical point $\delta=0$.

The Hamiltonian in eq(\ref{hamilton}) is closely related to the
transfer matrix of a classical two-dimensional layered Ising model,
which was first introduced and partially solved by McCoy and
Wu\cite{mccoywu}. Later the critical properties of the quantum model
was studied by Shankar and Murthy\cite{shankar}, and in great detail
by Fisher\cite{fisher}. Through a renormalization group (RG)
transformation Fisher has obtained many new results on static
quantities and equal time correlations, which are claimed to be exact
for large scales, i.e. in the vicinity of the critical point.  Many of
Fisher's results have been checked numerically\cite{youngrieger} and
in addition new results have been obtained about critical density
profiles\cite{profiles}, time-dependent critical
correlations\cite{riegerigloi} and various probability distributions
and scaling functions\cite{youngrieger,bigpaper}.  Later, using simple
expressions about the surface magnetization and the energy gap several
exact results have been derived by making use of a mathematical
analogy with surviving random walks\cite{bigpaper}, see also
\onlinecite{fisheryoung}.

In the Griffiths phase, where the RG results are restricted to the
immediate vicinity of the critical point, i.e. as $\delta \to 0$,
numerical investigations both on temperature dependent\cite{young}
(specific heat, susceptibility) and dynamical quantities (spin-spin
autocorrelations, distribution of the energy gap and
susceptibility)\cite{bigpaper,youngrieger} have lead to the
conclusion, that the behavior of all these quantities is a consequence
of Griffiths-McCoy singularities and can be characterized by a single
varying exponent $z(\delta)$ in (\ref{auto}). Very recently an
analytical expression for $z(\delta)$ has been derived\cite{diffusion}
by using an exact mapping\cite{itr} between the Hamiltonian in
(\ref{hamilton}) and the Fokker-Planck operator of a random walk in a
random environment. The dynamical exponent, which is given by the
positive root of the equation:
\be
\left[\left({J \over h}\right)^{1/z}\right]_{\rm av}=1\;,
\label{1/z}
\ee
generally depends both on $\delta$ and on the distributions $\pi(J)$ and
$\rho(J)$. However it becomes universal, i.e.
distribution independent, in the vicinity of the critical point 
when $z(\delta) \approx 1/(2 \delta),~|\delta| \ll 1$, in accordance
with the RG results\cite{fisher}. The numerical results obtained
about different singular quantities in the Griffiths phase
are all in agreement with the analytical formula in (\ref{1/z}) and
the observed small deviations are attributed to finite-size
corrections\cite{young,bigpaper}.

The singular quantities studied so far in the Griffiths phase are all
related to the scaling properties of the lowest energy gap, which
explains the observation why a single varying exponent is sufficient
to characterize the singularities of the different quantities.  There
are, however, other observables, which are expected to be singular
too, but not connected directly to the first gap. For example one
could consider the distribution of the second (or some higher) gap.
By similar reasons as for the first gap these higher excitations are
also expected to vanish in the thermodynamic limit and the
corresponding probability distributions are described by new exponents
for small values of the gaps.  As another example we consider the
connected transverse spin autocorrelation function $G_l^e(\tau)=
\langle\langle\sigma_l^z(0)\sigma_l^z(\tau)\rangle\rangle$.  In the
two-dimensional classical version of (\ref{hamilton}), the McCoy-Wu
model, this function corresponds to the energy-density correlation
function in the direction, where the disorder is correlated. Therefore
we adopt in the following this terminology and call $G_l^e(\tau)$ the
energy-density autocorrelation function. Since the inverse time scale
for these correlations is, as we shall see, determined by the second
gap, one expects that also $[G^e]_{\rm av}(\tau)$ has an algebraic
decay:
\be
[G^e]_{\rm av}(\tau) \sim \tau^{-\eta_e}\;,
\label{energy}
\ee
with an exponent $\eta_e$. Finally one should mention the non-linear
susceptibility whose distribution is expected to be described by a new
varying exponent.

In this paper we extend previous numerical work and study the scaling
behavior of the above mentioned singular quantities in the Griffiths
phase. We present a phenomenological scaling theory and we confront
its predictions by results of numerical calculations, based on the
free-fermion representation of the Hamiltonian in (\ref{hamilton}). We
show that the physical quantities we studied are characterized by
power-law singularities with varying critical exponents, the value of
those are connected to the dynamical exponent through scaling
relations.

Throughout the paper we use two types of random distributions. In the
symmetric binary distribution the couplings could take two values
$\lambda>1$ and $1/\lambda$ with the same probability, while the
transverse-field is constant:
\beqn
\pi(J) = {1\over 2}\left(\delta(J-\lambda)+\delta(J-\lambda^{-1})\right)\;,
\cr
\rho(h) = \delta(h-h_0)\;.
\label{binary}
\eeqn
At the critical point $h_0=1$, whereas in the Griffiths phase,
$1<h_0<\lambda$, the dynamical exponent from
(\ref{1/z}) is determined by the equation:
\be
h_0^{1/z}={\rm cosh}\left(\ln \lambda \over z \right)\;.
\label{1/zbin}
\ee
In the uniform
distribution both the couplings and the fields have rectangular
distributions:
\beqn
\pi(J)=\cases{1,&for $0<J<1$\cr
                0,&otherwise\cr}\;\cr
\rho(h)=\cases{h_0^{-1},&for $0<h<h_0$\cr
                0,&otherwise\cr}\;.
\label{uniform}
\eeqn
The critical point is also at $h_0=1$, whereas the dynamical exponent
is given by the solution of the equation:
\be
z \ln \left(1-z^{-2}\right)=-\ln h_0\;,
\label{1/zuni}
\ee
where the Griffiths phase now extends to $1<h_0<\infty$.

The structure of the paper is the following. In Section 2. we present
the free fermion description of various dynamical quantities.
Phenomenological and scaling considerations are given in Section 3, the
numerical results are presented in Section 4. Finally, we close the paper
with a Discussion.

\section{Free fermion description of dynamical quantities}

We consider the random transverse-field Ising model in
eq(\ref{hamilton}) on a finite chain of length $L$ with free
boundary conditions. The Hamiltonian in
eq(\ref{hamilton}) is mapped through a Jordan-Wigner transformation
and a following canonical transformation \cite{liebetal} into a free
fermion model:
\be
H=\sum_{q=1}^L \epsilon_q \left(\eta_q^+\eta_q-{1 \over 2}\right)\;,
\label{fermion}
\ee
in terms of the $\eta_q^+$ ($\eta_q$) fermion creation (annihilation)
operators. The energy of modes, $\epsilon_q$, is obtained through the
solution of an eigenvalue problem, which necessitates the
diagonalization of a $2L \times 2L$ tridiagonal matrix with
non-vanishing matrix-elements $T_{2i-1,2i}=T_{2i,2i-1}=h_i$,
$i=1,2,\dots,L$ and $T_{2i,2i+1}= T_{2i+1,2i}=J_i$, $i=1,2,\dots,L-1$,
and denote the components of the eigenvectors $V_q$ as
$V_q(2i-1)=-\phi_q(i)$ and $V_q(2i)=\psi_q(i)$, $i=1,2,\dots,L$, i.e.\
\end{multicols}
\widetext
\noindent\rule{20.5pc}{.1mm}\rule{.1mm}{2mm}\hfill
\be
T = \left(
\matrix{
 0  & h_1 &     &       &       &       &     \cr
h_1 &  0  & J_1 &       &       &       &     \cr
 0  & J_1 &  0  & h_2   &       &       &     \cr
    &     & h_2 &  0    &\ddots &       &     \cr
    &     &      &\ddots&\ddots &J_{L-1}&     \cr
    &     &      &      &J_{L-1}&   0   & h_L \cr
    &     &      &      &       &  h_L  &  0  \cr}
\right)\quad,\qquad
V_q = \left(\matrix{
-\Phi_q(1)\cr
 \Psi_q(1)\cr
-\Phi_q(2)\cr
 \vdots\cr
 \Psi_q(L-1)\cr
-\Phi_q(L)\cr
 \Psi_q(L)}
\right)\quad.
\ee
\hfill\rule[-2mm]{.1mm}{2mm}\rule{20.5pc}{.1mm}
\begin{multicols}{2} 
\narrowtext
\noindent 
We consider only the $\epsilon_q \ge 0$ part of the
spectrum\cite{igloiturban96}.

The local susceptibility $\chi_l$ at site $l$
is defined through the local magnetization $m_l$ as:
\be
\chi_l=\lim_{H_l \to 0} {\delta m_l \over \delta H_l}\;,
\label{locsusc}
\ee
where $H_l$ is the strength of the local longitudinal field, which
enters the Hamiltonian (\ref{hamilton}) via an additional term $H_l
\sigma_l^x$. $\chi_l$ can be expressed as:
\be
\chi_l=2 \sum_i{|\langle i|\sigma_l^x|0\rangle|^2\over E_i-E_0}\;,
\label{locsusc1}
\ee
where $|0\rangle$ and $|i\rangle$ denote the ground state and the $i$th
excited state of $H$ in (\ref{hamilton}) with energies $E_0$ and $E_i$,
respectively. For boundary spins one has the simple expression:
\be
\chi_1=2\sum_q{|\phi_q(1)|^2 \over \epsilon_q} \;.
\label{surfsusc}
\ee
Similarly, the local non-linear susceptibility is defined by:
\be
\chi^{\rm nl}_l=\lim_{H_l \to 0} {\delta^3 m_l \over \delta H_l^3}\;,
\label{nonlocsusc}
\ee
and can be expressed as:
\end{multicols}
\widetext
\noindent\rule{20.5pc}{.1mm}\rule{.1mm}{2mm}\hfill
\be
\chi^{\rm nl}_l=24 \left\{ \sum_{i,j,k}\langle 0|\sigma_l^x|i\rangle{1 \over E_i-E_0}
\langle i|\sigma_l^x|j\rangle{1 \over E_j-E_0}
\langle j|\sigma_l^x|k\rangle{1 \over E_k-E_0}\langle k|\sigma_l^x|0\rangle
+\sum_{i}\left({\langle i|\sigma_l^x|0\rangle \over E_i-E_0} \right)^2
\sum_j{|\langle j|\sigma_l^x|0\rangle|^2\over E_j-E_0}\right\}\;.
\label{nonlocsusc1}
\ee
It should be noted that it is {\it not} the first sum on the r.h.s.\ 
of (\ref{nonlocsusc1}) that gives the leading contribution, since at
least one of the 3 energy differences most involve a higher excitation
($\langle i|\sigma_l^x|j\rangle=0$ for $i=j$). For surface spins,
$l=1$, (\ref{nonlocsusc1}) simplifies:
\be
\chi^{\rm nl}_1=24\left\{ \sum_{p,q} {\phi_p(1)^2 \phi_q(1)^2 \over (\epsilon_p
+\epsilon_q)\epsilon_p}\left({1 \over \epsilon_p}-{1 \over \epsilon_q}
\right)-
\sum_p\left({\phi_p(1) \over \epsilon_p} \right)^2
\sum_q{|\phi_q(1)|^2 \over \epsilon_q}\right\} \;.
\label{nonsurfsusc}
\ee
\hfill\rule[-2mm]{.1mm}{2mm}\rule{20.5pc}{.1mm}
\begin{multicols}{2} 
\narrowtext
\noindent 
Next we consider the energy-density correlation function at site $l$, $G^e_l$,
defined by:
\beqn
G_l^{e}(\tau)=\langle 0|\sigma_l^{z}(\tau)\sigma_l^{z}(0)|0\rangle-
\langle 0|\sigma_l^{z}(\tau)|0\rangle\langle 0|\sigma_l^{z}(0)|0\rangle\cr
= \sum_{i>0}|\langle 0|\sigma_l^{z}|0\rangle|^2 \exp[-\tau(E_i-E_0)]\;.
\label{gmu}
\eeqn
In the free-fermion representation it is given by:
\be
G^e_l(\tau)=\sum_{\delta >\gamma} \left| \psi_{\delta}(l) \phi_{\gamma}(l)
-\psi_{\gamma}(l) \phi_{\delta}(l) \right|^2 \exp[-\tau(\epsilon_{\delta}
+\epsilon_{\gamma})]\;,
\label{zcorr}
\ee
which can be expressed for surface spins as:
\be
G^e_1(\tau)=\sum_{\delta >\gamma} \left[ {\epsilon_{\delta}-\epsilon_{\gamma}
\over h_1}
  \phi_{\delta}(1) \phi_{\gamma}(l) \right]^2 \exp[-\tau(\epsilon_{\delta}
+\epsilon_{\gamma})]\;.
\label{zsurfcorr}
\ee
The spin-spin autocorrelation function, $G_l$, which is defined as
$G_l^e$ in (\ref{gmu}) by replacing $\sigma_l^z$ by $\sigma_l^x$, is
generally complicated and can be expressed in the form of
Pfaffians\cite{stolze,bigpaper}. Exception is the autocorrelation
function for surface spins, which is simply given by:
\be
G_1(\tau)=\sum_q | \Phi_q(1)|^2 \exp(-\tau \epsilon_q)\;.
\label{xsurfcorr}
\ee

\section{Phenomenological and scaling considerations}

As described in the Introduction the Griffiths-McCoy singularities in
the paramagnetic phase are connected to the presence of strongly
coupled clusters, which are locally in the "ordered phase" and
therefore the corresponding excitation energy is very small. For the
RTIM the origin of these clusters can be explained either through the
analysis of the RG fixed-point distribution\cite{fisher}, which works
only in the vicinity of the critical point, or by using simple
explicit expressions for the excitation energy\cite{itks,bigpaper} and
estimate those through random walk arguments\cite{bigpaper}. Here we
use a simple phenomenological approach \cite{qsg,sglass,thill},
whose results are in agreement with the above microscopic methods.

Consider the quantity, $P_L(N)$, which measures the probability
that in a chain of $L$ sites there is a cluster of $N\ll L$ strongly
coupled spins. Since $N$ consecutive strong bonds can be found with exponentially
small probability $\sim \exp(-AN)$, whereas the cluster could be placed at
$\sim L$ different sites we have
\be
P_L(N) \sim L \exp(-AN)\;.
\label{PLN}
\ee
The excitation energy of this sample corresponds
to the energy needed to flip all spins in the cluster, which is
exponentially small in $N$:
\be
\epsilon_1 \sim \exp(-BN)\;.
\label{epsilon1}
\ee
Combining (\ref{PLN}) with (\ref{epsilon1}) we have for the probability
distribution of the first gap
\be
P_L(\ln \epsilon_1) \sim L \epsilon_1^{1/z}\;,
\label{PLeps1}
\ee
for $\epsilon_1 \to 0$ and $1/z=A/B$. Here, from the scaling combination
in eq(\ref{PLeps1}): $L \sim \epsilon_1^{-1/z} \sim \tau^{1/z}$, we can
identify $z$ as the dynamical exponent.

Next, we consider the second gap, $\epsilon_2$, which is connected to the
existence of a second strongly connected cluster of $N' \le N$ spins, and
its value corresponds to the energy needed to flip
all the spins in the second cluster simultaneously, consequently
\be
\epsilon_2 \sim \exp(-BN')\;.
\label{epsilon2}
\ee
The probability with which a cluster of size $N'$ occurs, provided
another cluster of size $N\ge N'$ exists, is given by $P'_L(N')\sim L
\exp(-AN') \sum_{N=N'}^L P_L(N)$. For $N'\ll L$ (or in the infinite
system size limit $L\to\infty$) this can be estimated as:
\be
P'_L(N') \sim L^2 \exp[-2AN']\;.
\label{PLNN}
\ee
Thus from (\ref{epsilon2}) and (\ref{PLNN}) we have
\be
P'_L(\ln \epsilon_2) \sim L^2 \epsilon_2^{1/z'}\;,
\label{PLeps2}
\ee
with $1/z'=2A/B$, thus
\be
z'=z/2\;.
\label{zprime}
\ee
Note that the scaling combination in the r.h.s. of (\ref{PLeps2}) is
dimensionless, as it should be.
Repeating the above argument for the
third, or generally the $n$th gap the corresponding distribution
is described by an exponent $z^{(n)}=z/n$, however the finite size
corrections for these gaps are expected to increase rapidly with $n$.

The scaling behavior of the probability distribution of the susceptibilities
can be obtained by noticing that both for $\chi_l$ and $\chi^{\rm nl}_l$ the
leading size dependence is connected with energy gaps in the numerator of
(\ref{locsusc1}) and (\ref{nonlocsusc1}), respectively. Then for the
asymptotic behavior of the distribution of the local susceptibility we
have:
\be
\ln\left[P(\ln \chi_l)\right] = -{1 \over z} \ln \chi_l + const\;,
\label{suscdist}
\ee
like to the inverse gap. For the non-linear susceptibility the second
term in the r.h.s. of Eq.(\ref{nonlocsusc1}) gives the singular contribution,
so that
\be
\ln\left[P(\ln \chi^{\rm nl}_l)\right] = -{1 \over z^{\rm nl}} \ln \chi^{\rm nl}_l + const\;,
\label{nonsuscdist}
\ee
with
\be
z^{\rm nl}=3z\;,
\label{znl}
\ee
since the asymptotic distribution is the same as that of the third
power of the inverse gap. We note that the relation in (\ref{znl})
corresponds to the phenomenological result in\cite{sglass}.

The scaling behavior of the average spin autocorrelation function is
given by:
\be
[G_l]_{\rm av}(\tau)=\int P_L(\epsilon_1) |M_l|^2 \exp(-\tau \epsilon_1) {\rm d}
\epsilon_1\;,
\label{intx}
\ee
where the factor with the matrix-element is $|M_l|^2 \sim 1/L$, since
the probability that a low energy cluster is localized at a given site,
$l$, is inversely proportional to the length of the chain. Then using
(\ref{PLeps1}) one arrives to the result in (\ref{auto}), thus establishing
the relation between the decay exponent of the spin autocorrelation
function and the dynamical exponent.

For energy-density autocorrelations, according to (\ref{zcorr}) and
(\ref{zsurfcorr}) the characteristic energy scale is $\epsilon_2$ and
the asymptotic behavior of the average energy-density autocorrelation function
is given by:
\be [G^e_l]_{\rm av}(\tau)=\int P'_L(\epsilon_2) |M_l^e|^2 \exp(-\tau
\epsilon_2) {\rm d} \epsilon_2\;.
\label{intz}
\ee
Now we take the example of the surface autocorrelation function in
(\ref{zsurfcorr}) to show that the factor with the matrix-element,
$|M_1^e|^2$, is proportional to $\epsilon_2^2$. The remaining factor
in (\ref{zsurfcorr}) with the first components of the eigenvectors is
expected to scale as $1/L$ due to similar reasons as for the spin
autocorrelations, thus $|M_l^e|^2\sim L^{-1} \epsilon_2^2$ and
together with (\ref{PLeps2}) one has $P'_L(\epsilon_2) |M_l^e|^2 \sim
L \epsilon_2^{1/z'+1}$.  Before evaluating the integral in
(\ref{intz}) we note that for a fixed $L$ the expression in
(\ref{intz}) stays valid up to $\tau \sim L^z$. Therefore to obtain
the $L$ independent asymptotic behavior in $\tau$ we should instead
vary $L$, so that according to (\ref{PLeps2}) take $L \sim
\epsilon_2^{-1/(2z')}$ and in this way we stay within the border of
validity of (\ref{intz}) for any $\tau$. With this modification we
arrive to the result in Eq.(\ref{energy}) where the decay exponent,
$\eta_e$, is related to the dynamical exponent as
\be
\eta_e=2+{1 \over z}\;,
\label{etaz}
\ee
where the relation in (\ref{zprime}) is used.  We expect that the
factor, $|M_l^e|^2$, has the same type of scaling behavior for any
position $l$, thus the relation in (\ref{etaz}) stays valid both for
bulk and surface spins. We note that the reasoning above (\ref{etaz})
applies also for the spin autocorrelation function, in which case in
(\ref{intx}), however there is no explicit $L$ dependence.

In this way we have established a phenomenological, scaling theory which makes a connection
between the unconventional exponents in (\ref{zprime}), (\ref{znl})
(\ref{etaz}) and the dynamical
exponent. In the next Section we confront these relations with numerical
results.

\section{Numerical results}

In the numerical calculations we have considered RTIM chains with
up to $L=128$ sites and the average is performed over several 10000
realizations, typically we considered 50000 samples. For some cases,
where the finite size corrections were strong, we also made runs with
$L=256$, but with somewhat less realizations. 

We start to present results about the distribution of the first and
second gaps. As illustrated in Fig. 1, both for the uniform and the binary
distributions, the asymptotic scaling relations for the distribution of
the first two gaps in (\ref{PLeps1}) and (\ref{PLeps2}) are satisfied. From
the asymptotic slopes of the distributions in Fig. 1 we have estimated the
$1/z$ and $1/z'$ exponents for the
two largest finite systems, $L=64$ and $L=128$, which are presented in
Fig. 2  for different points of
the Griffiths phase for the uniform distribution. As seen in the Figure
the $z$ exponent calculated from the first gap agrees very
well with the analytical results in (\ref{1/zuni}).
For the $z'$ exponent, as calculated from the distribution of the second gap
the scaling result in (\ref{zprime}) is also well satisfied, although the
finite-size corrections are stronger than for the first gap.
For the third gap, due to the even stronger finite size effects, we have
not made a detailed investigation. Extrapolated results at $h_0=2$
are found to follow the scaling result $z^{(3)}=z/3$.

Next, we study distribution of the linear and non-linear local
susceptibilities at the surface spin. As demonstrated in Fig. 3 both type
of distributions satisfy the respective asymptotic relations in
(\ref{suscdist}) and (\ref{nonsuscdist}), from which the critical exponents
$z$ and $z^{\rm nl}$ are calculated. The estimates are shown in Fig. 4 at different
points of the Griffiths phase. As seen in the Figure the numerical results
for the dynamical exponent, $z$, are again in very good agreement with the
analytical results in (\ref{1/zuni}) and also the exponent of the non-linear
susceptibility, $z^{\rm nl}$, follows fairly well the
scaling relation in (\ref{znl}).

Finally, we calculate the average energy-density autocorrelation
function. As seen in Fig. 5 $[G^e]_{\rm av}(\tau)$ displays a linear
region in a log-log plot, the size of which is increasing with $L$,
but its slope, which is just the decay exponent, $\eta_e$, has only a
weak $L$ dependence.  The slope of the curve and thus the
corresponding decay exponent $\eta_e$ has a variation with the
parameter $h_0$, as illustrated in Fig. 6. The estimated $\eta_e$
exponents at the critical point, $h_0=1$, and in the Griffiths phase
are presented in Fig. 7. As seen in this Figure the variation of
$\eta_e$ is well described by the form $\eta_e(\delta)=
\eta_e(0)+1/z(\delta)$. This functional form corresponds to the
scaling result in Eq.(\ref{etaz}), if the critical point correlations
decay with
\be
\eta_e(0)=2\;.
\label{etaz0}
\ee
The numerical calculations with $L=128$ give a
slightly higher value $\eta_e(0) \approx 2.2$ \cite{riegerigloi}. However,
the finite-size estimates show a slowly decreasing $\eta_e(0)$ with increasing
system size. Repeating the calculation with $L=256$ we got $\eta_e(0) \approx
2.1$. Thus we can conclude that the scaling relation in (\ref{etaz}) is
probably valid and then Eq.(\ref{etaz0}) is the exact value of the decay
exponent of the average critical energy-density autocorrelations\cite{surface}.

\section{Discussion}

In this paper we have considered the random transverse-field Ising
spin chain and studied different consequences of the Griffiths-McCoy
singularities in the paramagnetic phase. Our main conclusion is that
all singular quantities can be characterized by power-law
singularities and the corresponding varying critical exponents can be
related to the $z(\delta)$ dynamical exponent and, for energy-density
autocorrelations, to the $\eta_e(0)$ critical point exponent. Since
the exact value of $z(\delta)$ is known in (\ref{1/z}) and we expect
that also the relation in (\ref{etaz0}) is valid, thus we have a
complete, analytical description of the Griffiths phase of the RTIM in
one dimension.

One interesting feature of our results concerns the distribution of
the higher excitations and the value of the corresponding exponent
$z^{(n)}=z/n$. Since the decay of dynamical correlations of general, more
complex operators are related to $1/z^{(n)}=n/z$, we obtain a hierarchy
of decay exponents which could be simply expressed by those of a few
primary operators. This feature is reminiscent to the tower-like structure
of anomalous dimensions in two-dimensional conformal models\cite{cardy}. Our
knowledge about the higher excitations can also be used to estimate the
correction to scaling contributions.

Much of the reasoning of our phenomenological scaling considerations
in Section 3 stay valid for other random quantum systems. Especially
the scaling behavior of the higher gaps and the corresponding relation
in Eq.(\ref{zprime}) should be valid even for higher dimensions and
the same is true for the distribution of the non-linear susceptibility
and the corresponding relation in (\ref{znl}).

In one dimensions the universality class of the RTIM involves several
random systems, among others the random quantum Potts
chain\cite{potts}. For these models one does not expect a universality
of the $z(\delta)$ exponent in the Griffiths phase, although scaling
relations, like the one in (\ref{zprime}) are very probably valid. It
would be interesting to perform a numerical study on the random
quantum Potts model to check the existing conjectures.

Another possible field where the present results could be applied is
the problem of anomalous diffusion in a random environment
\cite{diffusion,itr,diffother}.  Making use of the exact
correspondence\cite{diffusion,itr} between the Hamilton operator in
(\ref{hamilton}) and the Fokker-Planck operator of the one-dimensional
random walk we can use the relation in (\ref{zprime}) to describe the
distribution of the eigenvalues of the corresponding Fokker-Planck
operator. One can also define analogous quantity to the energy-density
autocorrelation function in (\ref{zcorr}) by considering connected
persistence correlations, whose asymptotic decay is related to the
distribution of the second gap, like in (\ref{intz}). Research in this
field is in progress.

Acknowledgment: This study has been partially performed during our
visits in K\"oln and Budapest, respectively. F.\ I.'s work has been
supported by the Hungarian National Research Fund under grant No OTKA
TO23642 and by the Ministery of Education under grant No. FKFP 0765/1997.
H.\ R. was supported by the Deutsche Forschungsgemeinschaft (DFG).

\begin{figure}
\epsfxsize=\columnwidth\epsfbox{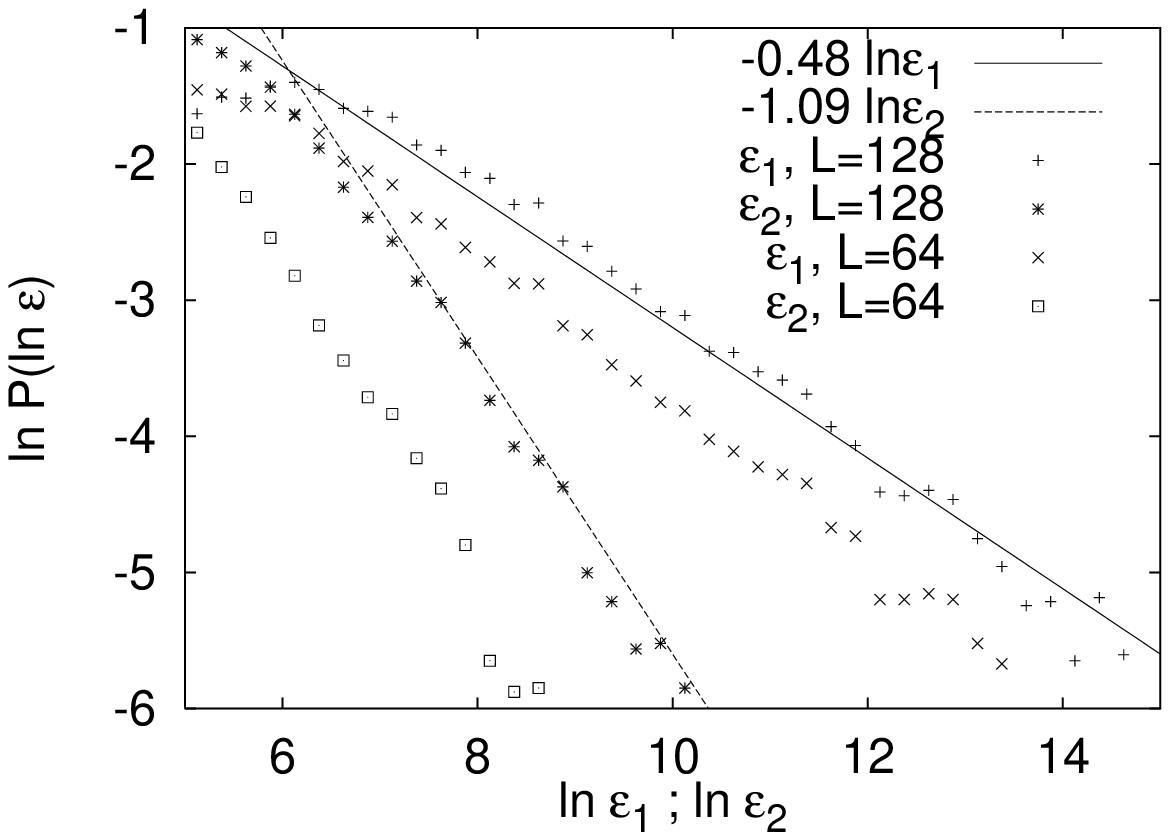}
\epsfxsize=\columnwidth\epsfbox{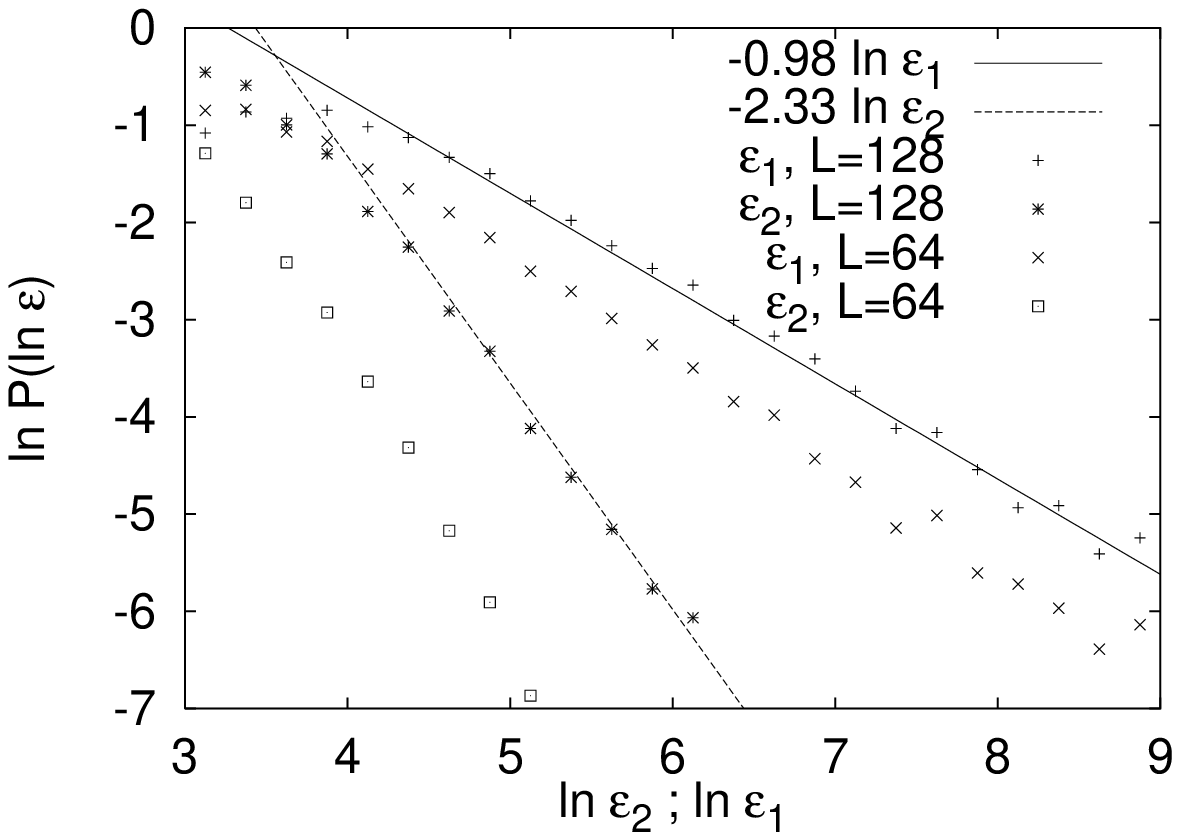}
\caption{
  Probability distribution of $\ln\varepsilon_1$ and
  $\ln\varepsilon_2$ for the uniform distribution at $h_0=2$ (top) and
  the binary distribution ($\lambda=4$) at $h_0=2.5$ (bottom).  The
  straight lines are least square fits to the data for the largest
  system size, their slopes correspond to $1/z(h_0)$ and $1/z'(h_0)$,
  respectively. They follow the predicted relation $z'(h_0)=z(h_0)/2$.
\label{fig1}
}
\end{figure}

\begin{figure}
\epsfxsize=\columnwidth\epsfbox{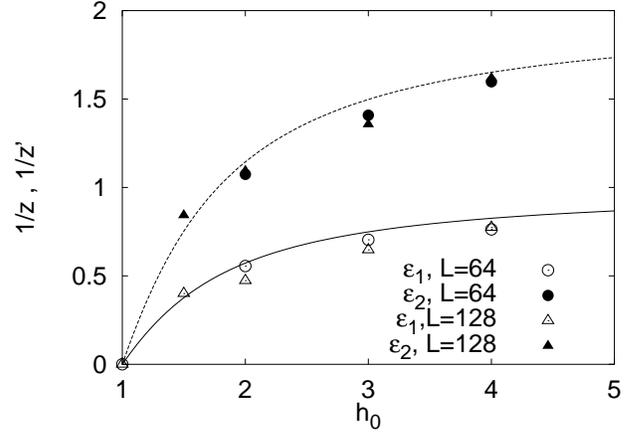}
\caption{
  The estimates for $1/z$ and $1/z'$ as a function of $h_0$ for the
  uniform distribution. They have been obtained from our analysis of
  the probability distribution of $\ln\varepsilon_1$ and
  $\ln\varepsilon_2$, respectively, for two system sizes (as
  exemplified in fig. 1). The full line for $1/z$ corresponds to the
  analytical result (\protect{\ref{1/z}}), the broken line corresponds
  to $2/z$, which we predict to be identical with $1/z'$.
\label{fig2}
}
\end{figure}

\begin{figure}
\epsfxsize=\columnwidth\epsfbox{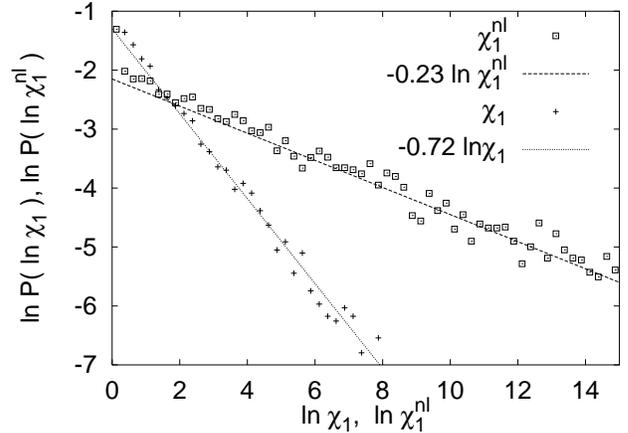}
\caption{
  Probability distribution of the linear and non-linear
  susceptibility, $\ln\chi_1$ and $\ln\chi_1^{\rm nl}$, respectively,
  for the uniform distribution at $h_0=3$. The straight lines are
  least square fits to the data for the largest system size, their
  slopes correspond to $1/z(h_0)$ and $1/z^{\rm nl}(h_0)$,
  respectively. They follow the predicted relation $z^{\rm
    nl}(h_0)=3z(h_0)$.
\label{fig3}
}
\end{figure}

\begin{figure}
\epsfxsize=\columnwidth\epsfbox{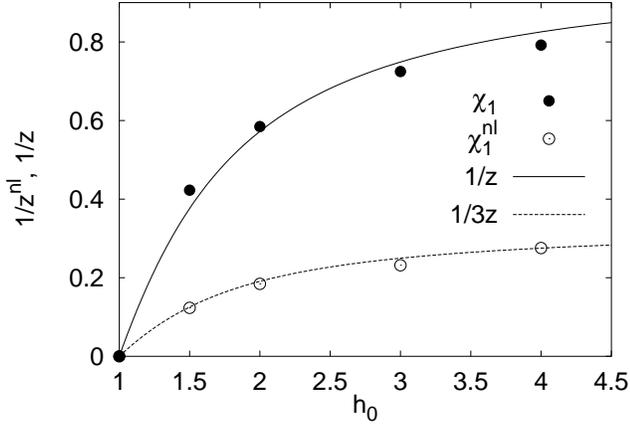}
\caption{
  The estimates for $1/z$ and $1/z^{\rm nl}$ as a function of $h_0$
  for the uniform distribution. They have been obtained from our
  analysis of the probability distribution of $\ln\chi_1$ and
  $\ln\chi_1^{\rm nl}$, respectively, for two system sizes (as
  exemplified in fig. 3). The full line for $1/z$ corresponds to the
  analytical result (\protect{\ref{1/z}}), the broken line corresponds
  to $1/3z$, which should be identical with $1/z^{\rm nl}$.
\label{fig4}
}
\end{figure}

\begin{figure}
\epsfxsize=\columnwidth\epsfbox{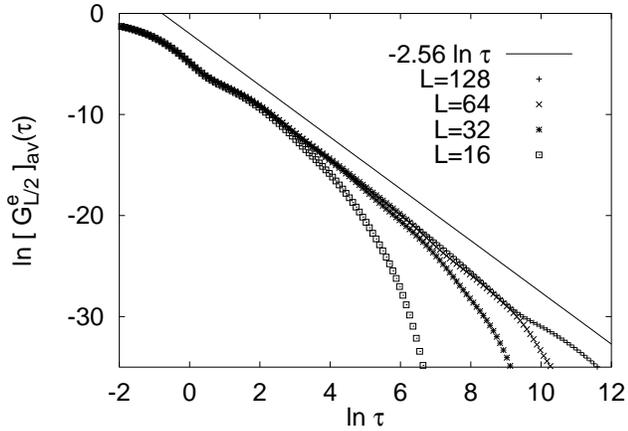}
\caption{
  The bulk energy-energy autocorrelation function $[G_{L/2}^e]_{\rm
    av}(\tau)$ for the binary distribution ($\lambda=4$) at $h_0=1.5$
  for different system sizes as a function of $\ln\tau$. The slope of
  the straight line identifies the exponent $\eta_e(h_0)$ describing
  the asymptotic decay of $[G_{L/2}^e]_{\rm av}(\tau)$ in the infinite
  system size limit $L\to\infty$.
\label{fig5}
}
\end{figure}

\begin{figure}
\epsfxsize=\columnwidth\epsfbox{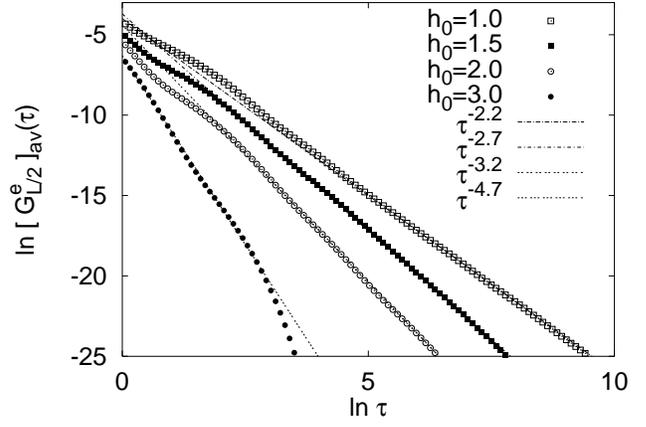}
\caption{
  The bulk energy-energy autocorrelation function
  $[G_{L/2}^e]_{\rm av}(\tau)$ for the binary distribution ($\lambda=4$) at
  different values for $h_0$ for $L=128$ as a function of $\ln\tau$.
  One observes the variation of the exponent $\eta_e(h_0)$ (identical
  to the slope of the straight line fits) with increasing $h_0$.
\label{fig6}
}
\end{figure}

\begin{figure}
\epsfxsize=\columnwidth\epsfbox{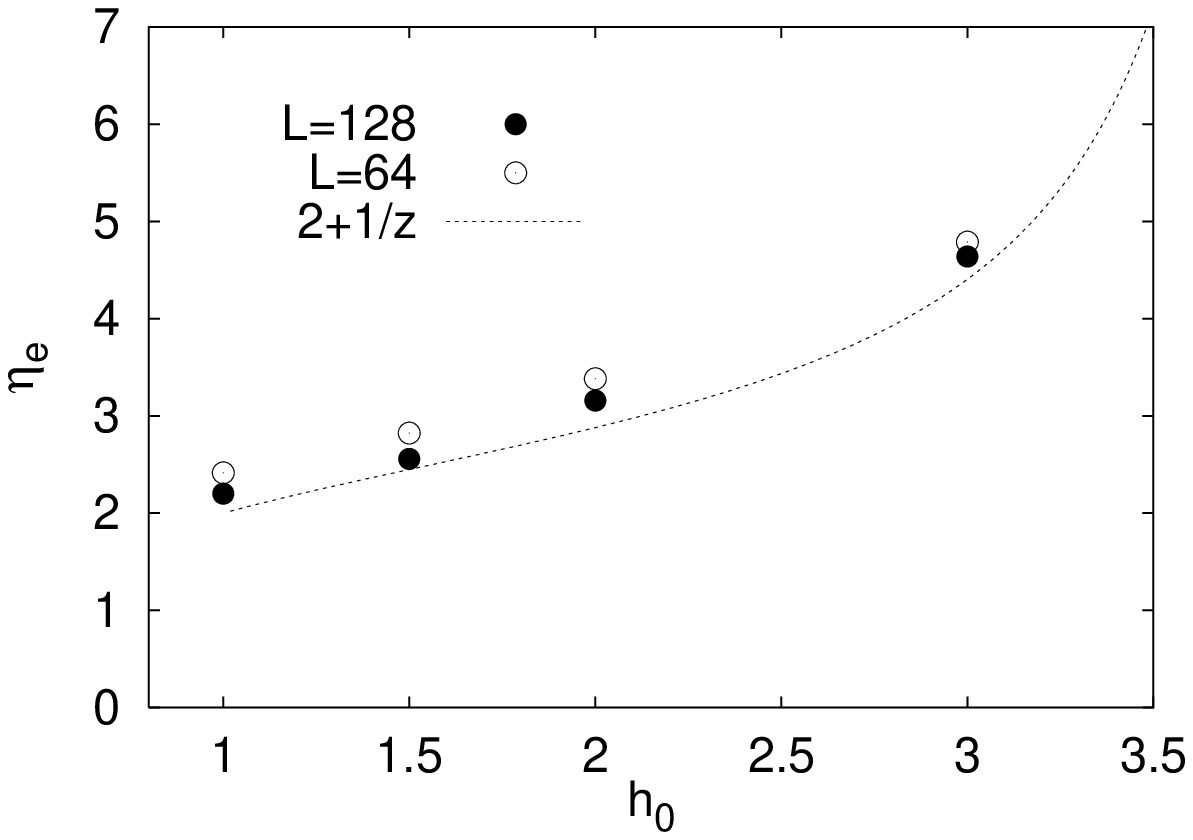}
\caption{
  The exponent $\eta_e(h_0)$ for the binary distribution ($\lambda=4$)
  as obtained from the analysis of the asymptotic decay of the bulk
  energy-energy autocorrelation function $[G_{L/2}^e]_{\rm av}(\tau)$
  \'a la fig. 6. The full line is the analytical prediction
  $\eta_e(h_0)=2+1/z(h_0)$ with $z(h_0)$ given by the exact formula
  (\protect{\ref{1/z}}) for the binary distribution with $\lambda=4$.
\label{fig7}
}
\end{figure}

\end{multicols}


\begin{references}
\bibitem{classical}
        In a classical random system the singularities in the
        corresponding Griffiths-region are weaker, e.g. the
        spin-spin autocorrelation function has an enhanced power law
        decay (i.e.\ $\sim\exp(-c(\ln t)^\alpha)$ with $\alpha>1$),
        , in contrast to the power-law dependence observed in
        quantum systems.
\bibitem{surface}
        Numerical estimates for the decay exponent of the average
        energy-density autocorrelation function for surface spins at
        the critical point are $\eta_e^s \approx 2.5$ with
        $L=128$\cite{riegerigloi}, which is somewhat larger than for
        bulk autocorrelations. Discrepancies between estimates for $z$ 
        from surface and bulk quantities have been observed 
        before\cite{bigpaper}. They can be attributed to corrections 
        to scaling effects which are different for different 
        quantities, see also fig.6 in \cite{young}.
\end{references}
\end{document}